# Stress-transfer from polymer substrates to monolayer and few-layer graphenes

Ch. Androulidakis[1], D. Sourlantzis[2], E.N. Koukaras[1,3], A.C. Manikas[2] and C. Galiotis[1,2]*

[1]Institute of Chemical Engineering Sciences, Foundation of Research and Technology-Hellas (FORTH/ICE-HT), Stadiou Street, Platani, Patras, 26504 Greece
[2]Department of Chemical Engineering, University of Patras, Patras, 26504 Greece
[3]School of Science & Technology, Hellenic Open University, Patras, 26222 Greece

*Corresponding author: c.galiotis@iceht.forth.gr, galiotis@chemeng.upatras.gr

## ABSTRACT

In the present study the stress transfer mechanism in graphene-polymer systems under tension is examined experimentally using the technique of laser Raman microscopy. We discuss in detail the effect of graphene edge geometry, lateral size and thickness which need to be taken under consideration when using graphene as a protective layer. The systems examined comprised of graphene flakes with large length (over ~50 microns) and thickness of one to three layers simply-deposited onto PMMA substrates which were then loaded to tension up to ~1.60% strain. The stress transfer profiles were found to be linear while the results show that large lateral sizes of over twenty microns are needed in order to provide effective reinforcement at levels of strain higher than 1%. Moreover, the stress-built up has been found to be quite sensitive to both edge shape and geometry of the loaded flake. Finally, the transfer lengths were found to increase with the increase of graphene layers. The outcomes of the present study provide crucial insight on the issue of stress transfer from polymer to nano-inclusions as a function of edge geometry, lateral size and thickness in a number of applications.

**1. Introduction**

Graphitic materials in various forms such as carbon fibres (CFs), carbon nanotubes (CNTs) and graphene (Gr) exhibit remarkable mechanical properties, such as high moduli up to ~1 TPa[1-4] and tensile strengths up to ~100 GPa[1], which can be put into good use in composites employing polymer[2], ceramic[3] and even metal matrices[4]. In the case of CF/ polymer composites there is already a dearth of commercial applications in the aerospace, automotive, household, and recreational sectors that have sprung out over the last 30 years due to the unique combination of mechanical property per unit mass that they can offer. CNTs and graphene have also been examined recently as nano-fillers in polymer matrices as they can provide moderate enhancement in modulus and strength at small loadings in combination with a significant increase of thermal and electrical conductivities of the host matrices[5-6]. Graphene in particular, offers certain advantages over CNTs as it can be handled much more easily and its high surface area makes it more effective as a potential filler for engineering polymers[7-8]. Moreover, recent studies have shown that single and few layer graphenes are very effective for reinforcing metals such as nickel and palladium due to the strong interfacial bonding that is developed between these two classes of materials[4, 9]. Besides its use as reinforcing filler, graphene is used as a coating in conventional materials for inducing multi-functionality[10]. Various graphene coated applications have been demonstrated such as gilding of large structures[4] and protection from corrosion[11].

For producing graphene/ polymer composites even at low volume fractions, methods for scalable synthesis of graphene, such as shear liquid exfoliation[12-13], that yield relatively large quantities of multi-layer graphene flakes need to be employed. In fact, in a recent paper[14], it has been shown that the commercially available graphene with scalable production consists of few-layers of lateral size of only a few microns. Furthermore, in most cases the resulting graphene flakes are of small size (~3-5 μm) and of irregular shape. It is therefore clear that both size, thickness and shape of the flakes employed can play a critical role in the reinforcing capabilities of the filler[15-16]. Despite the extensive use of these few-layer graphenes in nano-composites there is very limited experimental work on the corresponding stress transfer mechanisms[17] and particularly the required characteristics for efficient reinforcing capabilities.

The effective use and design of graphene as reinforcing agent lies on the understanding of the interfacial behaviour of the graphene/matrix system. The mechanical load is transferred from the polymer to the graphitic inclusion by the interfacial shear, which is described by the well-known shear- lag mechanism in composite materials[18-19]. The stress is built from the edge and increases towards the inner part until reaches the maximum value at some distance away from the edge which is usually termed as the transfer length ($l_t$) (**figure 1**). The critical flake length ($l_c$) will then be the minimum value required for the total stress build up and is defined as $l_c= 2l_t$.

The stress transfer mechanism of single layer graphene simply supported on polymers like PMMA[20] and PET[21-22] under tension has already been examined. The transfer length from the previous studies was found to be in the range of 4-10 microns[16-18]. These values obtained from flakes of relatively small length (~15 μm or less), and as will be shown later on this is not the actual case since the transfer length is strain dependent. The shear-lag effect holds for small strain levels after which strain is transferred through friction with a constant interface shear stress (ISS), until interface failure results in sliding of the graphene and no further transfer of mechanical stress is then possible[22]. The ISS has been reported to be in the range of 0.40-0.80 MPa for graphene-polymer systems by converting the Raman data to strain maps and by considering the balance of shear-to-axial forces at the interface[17, 20, 22-23]. These values are relatively low and methods for increase them have been proposed by previous work of the group, such as the creation of a wrinkled interface which significantly enhances the stress transfer efficiency from the polymer to graphene[16] or the introduction of artificial 'defects' to the filler that enhances the anchoring of graphene to the host polymer[24]. Moreover, the chemical modification of the interface polymer-graphene has proven to be effective in increasing the ISS[25].

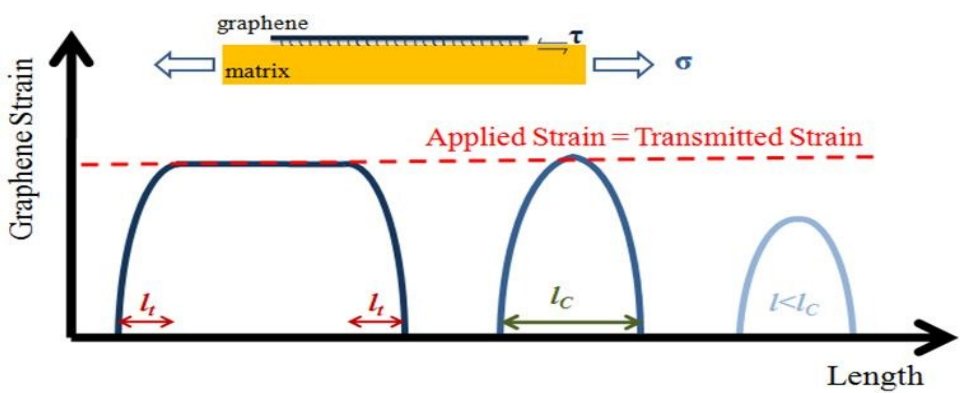

**Figure 1.** Schematic of the stress-transfer shear mechanism in graphene/polymer system. For large length of graphene the applied strain to the matrix is fully transmitted to the graphene. The length required for the strain build-up is the transfer length ($l_t$) and required critical length ($l_c$) of the flake for efficient reinforcement is $l_c=2l_t$.

In order to estimate the values of ISS, the strain profile and build-up from the edges of the graphene towards the inner part needs to be captured as seen in **figure 1**[20, 22]. In this regard Raman spectroscopy has proven to be the most efficient method; the Raman peaks shift with the application of mechanical load[26] and by monitoring the shift rates of the peaks (2D[27] and G[28]), the level of stain in the graphene can be back-tracked in a straightforward manner[20, 22]. Raman maps can be taken for the 2D and G phonons across graphene flakes at small steps, even at the sub-micron level, providing high resolution for the strain distribution in graphene[20]. The Raman maps are converted to values of strain based on the average shift rate which has been examined in numerous works to be in the range of ~50-64 cm$^{-1}$/% for the 2D peak[20, 27, 29-30]. As mentioned earlier, having established the strain build-up, the ISS can be obtained by a simple balance-of-axial-to-shear-forces argument[20].

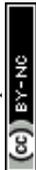

Despite the recent extensive efforts to exploit graphene as a reinforcing filler in polymers, the stress transfer characteristics have been studied in a handful of studies only on monolayer graphene/ polymer systems whereas studies of few-layer graphene deposited or embedded in polymers are scarce[17, 31]. This is indeed a problem for engineering applications of graphene composites as most mass exfoliation techniques yield distributions of flakes of various thicknesses and lateral sizes. Moreover, as mentioned and above graphene is used as a coating material for large surfaces and thus, it is of crucial importance a comprehensive investigation of interfacial interactions with graphene of various thicknesses on substrate. Herein, we examine the stress-transfer mechanism of large graphene flakes with length of >50 microns for thickness of one to three layers simply supported on engineering polymers, such as PMMA, using the methodology presented above. Various aspects that have not been received attention in previous studies such as the effect of the graphene geometry on the ISS, and the implications of the stress transfer through friction on the transfer length are highlighted. The present work provides significant guidelines and in-depth understanding for the effective use of graphene in strain engineering applications and as protective coating.

## 2. Experimental section

Graphene flakes prepared in a clean room by mechanical exfoliation of highly ordered pyrolytic graphite (HOPG) using the scotch tape method[32]. The exfoliated graphitic materials deposited directly on the surface of the SU-8/ PMMA substrate. In **figure 2** the optical images of the examined graphene flakes are presented. The SU-8 photoresist [Microchem 2000.5] was spin coated on the surface of the PMMA bar with a speed of ~4000 rpm, resulting in a very thin layer of thickness ~180-200 nm. Curing of the SU-8 followed by three steps; pre-bake, UV exposed and post-bake treatment. Appropriate graphene flakes were located with an optical microscope and the exact thickness identified from the line-shape of the 2D Raman peak. A four-point-bending apparatus adjusted under the Raman microscope for simultaneously loading the specimens under tension and for recording the Raman spectra (785 nm excitation). The laser power was kept below ~1 mW to avoid local heating of the specimens. External strain was applied in a stepwise manner by bending of the polymer bar at increments of ~0.1% and the 2D Raman peak was acquired *in situ*. At every strain level the whole graphene flake was scanned from edge to edge at steps of ~1 μm and Raman spectra were continuously taken. The magnitude of strain on the top surface of the PMMA bar where graphene was located was estimated by the beam deflection and also by means of electrical resistance strain gauges[15-16, 20].

## 3. Results and Discussion

In **figure 2a** an optical image of the examined single layer graphene is presented. The flake is relatively large with a length of ~60 microns and has a micro-ribbon geometry. It is important to note that the right edge is almost perfectly vertically and normal to the tensile direction while the left edge is tapered at an angle of ~32 degrees with respect to the horizontal axis (**figure 2a**).

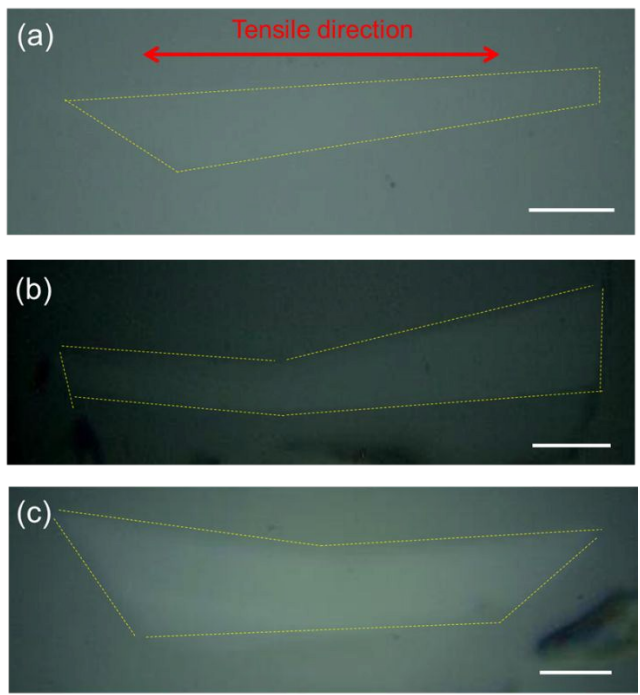

**Figure 2.** Optical images of the examined graphene flakes of thickness (a) one- layer (b), two-layers and (c) three-layers. All the flakes are relatively large and with ribbon type geometry. The yellow dotted lines denote the shape of the flakes and the geometry of the edges with respect to the tensile direction. The scale bar is 10 microns.

To measure accurately the Raman wavenumber shift for the monolayer graphene we ensure that the data were collected from the middle of the flake for which the stress/ strain distribution vs distance from the edges forms a plateau. Evidently measurements on small flakes can be problematic as the lateral size maybe smaller than the critical length (see further comments below). Here, the specimen was subjected to tension up to ~1.60% of strain and the average shift of the 2D peak was found to be ~−47.4±2 cm$^{−1}$/%, which is somewhat lower from the expected value of ~−55 cm$^{−1}$/% for a laser line of 785 nm[27]. The actual strain in the graphene can be obtained by the following equation[20]:

$$(1) \quad \varepsilon = \frac{\omega_{2D} - \omega_{2D,0}}{k_{2D}}$$

where $\omega_{2D}$ and $\omega_{2D,0}$ are the frequency of the 2D peak at every measured strain level and at rest, respectively, and $k_{2D}$ is the shift rate for the 785 nm laser line with value ~55 cm$^{-1}$/% as confirmed previously[27, 30, 33]. The strain profiles for various levels of tension are obtained by converting the 2D Raman maps taken across a line parallel to the direction of tension to values of strain using the above equation as seen in **figure 3b.** The flake is under residual tension at rest, and the build-up of the strain takes place from the edges towards the centre of the flake. By employing a balance-of-forces argument for the stress transfer in such model systems, the interfacial shear stress (ISS) is derived from equation 2[16, 20] and its derivation is presented in the SI:

$$(2) \quad \left(\frac{\partial \varepsilon}{\partial x}\right)_{T=298K} = -\frac{\tau_t}{nt_g E} \Leftrightarrow \tau_t = -nt_g E \left(\frac{\partial \varepsilon}{\partial x}\right)_{T=298K}$$

Where $\varepsilon$ is the applied strain, $\tau_t$ is the ISS, $E$ is the Young's modulus of monolayer graphene (~1 TPa), $n$ is the number of layers of the graphene and $t_g$ is the thickness of a single layer graphene (0.34 nm). The slopes $d\varepsilon/dx$ can be extracted from the Raman data and therefore the values of the interfacial shear stress per strain level are easily obtained (**figures 3-5**).

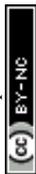
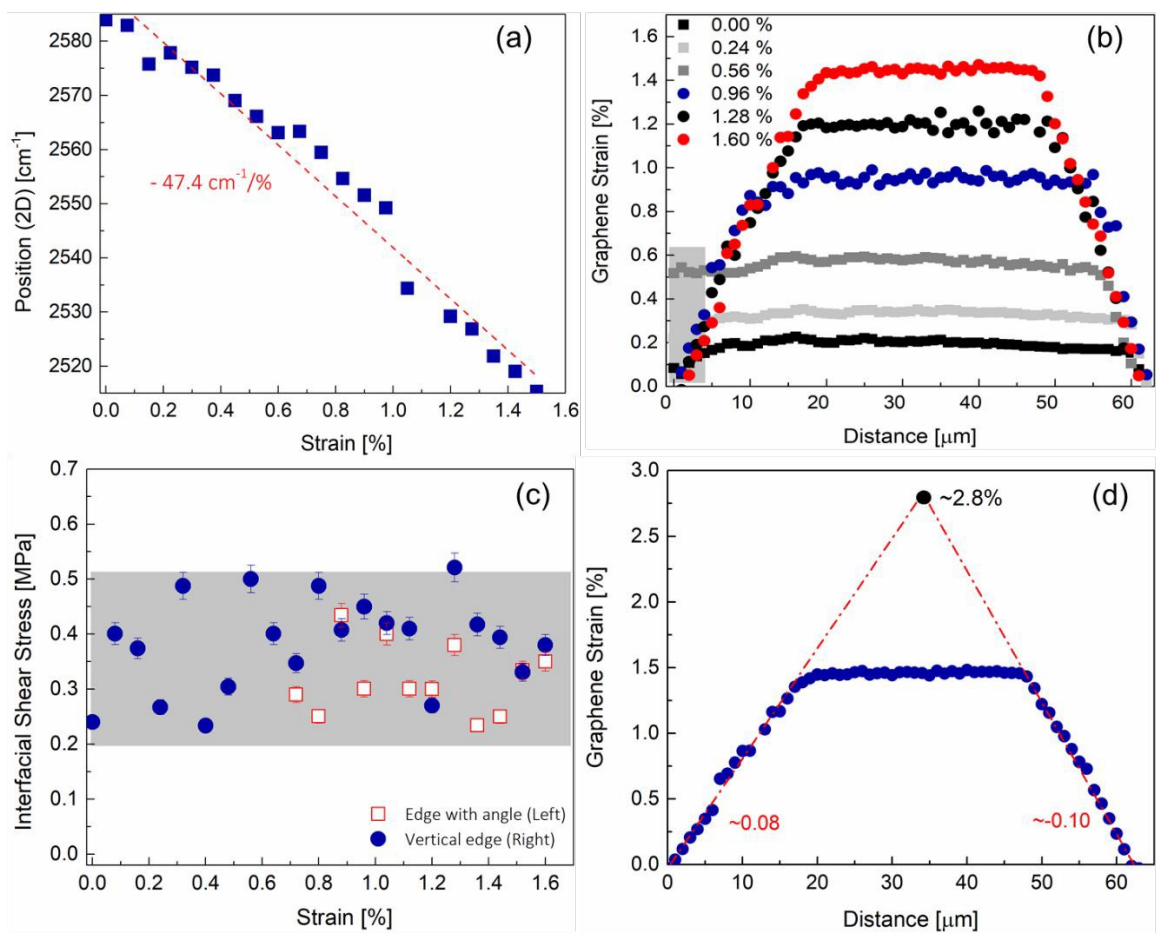

**Figure 3.** (a) Average shift of the 2D peak of the single layer graphene of the specimen examined here. (b) Representative strain distributions across the length of graphene for selected levels of strain. The strain derived by converting the spectroscopic data to values of strain. Until 0.56% the left angle indicates the presence of axial strain transmission as noted by the shaded grey area. (c) The interfacial shear stress (ISS) estimated from equation 3 for the left (angular) and right (aligned normal to the tensile direction) edges of the flake by linear fit of the strain build-up. (d) Expanding the fitting lines of the strain build-up until convergence. The top value of the formation of the triangle corresponds to the maximum strain that the graphene can reach before slipping from the substrate occurs. It is interesting to note that no fracture is observed up to 1.6% strain which was the upper limit of strain imparted to graphene by the experimental setup.

We can observe that the strain build-up is not the same for both edges, which indicates that the edge shape affects the ISS values. This is a significant point which has not received significant attention in previous works. The build-up from the square edge normal to applied stress follows the shearing mechanism and can be approximated with very good accuracy with linear fits, which indicates that friction like mechanism is prevalent in the as-supported graphene/ PMMA composite. It is interesting to note that very similar strain transfer profiles have been obtained in composite systems, such as in fibre/ metal composites, for which the fibres are weakly bonded to the matrix[34]. Furthermore, past work of the group has shown that linear profiles are also observed in untreated (and unsized) carbon fibre/ epoxy systems[35] whereas, in contrast, oxidised carbon fibres embedded in epoxy matrices exhibit quite pronounced shear lag behaviour in which the absolute value of the strain derivative of equation (2) in the transfer length region is not a constant value but decays abruptly as one moves towards the centre of the fibre[35-36]. Furthermore, the clear build-up from the very edges indicates no interface slippage between graphene and polymer, since the Raman peak positions shift with the applied strain.

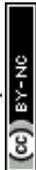

The strain built up profiles from the angular edge also shows a quite complex behaviour. For small strains a build-up from the edge toward the middle of the flake is observed but for higher strains it seems that there is an axial transmission of stress (strain) rather than a shear generated stress (strain) transfer (**figure 1b**). The reasons for the axial transmission are not quite clear but it is not inconceivable that the angular edge has been embedded or bonded into the resin due to the transfer process allowing axial transmission. From the Raman shift we measure a maximum stress of 5 GPa for applied strain of 0.56% and considering the angle of the edge we get ~4.3 GPa acting on the normal face of graphene. We must note that this is a rough estimation since the distribution of the normal stress is not constant throughout the edge area and changes with the applied strain as seen in **figure 3b**. This effect has been shown and examined elsewhere in detail by FEM simulations in similar system[37]. At applied strains of ~1% or even higher the axial transmission is lost possibly to the debonding of flake angular edge and then a shear mechanism prevails. The fact that the graphene strain now builds in a linear fashion from a zero value at the edge confirms the above assertion. Finally, it is worth noting that no fracture or interface failure is observed up to 1.60% strain which was the upper limit imparted to graphene by the experimental setup. This is quite significant since the overall loaded area is of the order of ~60 × 8= 480 μm² which is much larger than any such experiments reported earlier by us[20] or others[23] and shows once more than the crystal perfection of graphene is retained at much higher length scales than envisaged earlier.

In **figure 4a** the ISS profile across the length of the single layer graphene for the maximum tensile strain is presented and for various selected levels of strain are given in SI. The maximum ISS occurs at the very edges of the graphene flakes, which indicates the absence of doping at the edges[20]. As expected the ISS values (particularly in the case of the 'square' end) are of constant value and of opposite sign from each flake edge. For increasing tension, the length required for reaching the zero ISS increases too, and for the 'square' end is about ~10 μm at the maximum tensile applied strain of 1.6%. All ISS values from both ends (for the cases of shear transfer) are presented in **figure 3c**. As seen, the ISS values as a function of applied strain form a plateau - within the experimental error- with a mean value of ~0.39 MPa for the square end and ~0.31 MPa for the angular edge, respectively. The average ISS and the shift rate obtained herein are somewhat lower than the previous reported values of 0.45 MPa[20] and ~−55 cm$^{-1}$/%[27] for the same polymer substrate and for same preparation procedure. This small deviation is within the experimental error and thus acceptable, and stems from slight differences in the experimental conditions such as sample preparation and mechanical experiments. As is therefore evident the shape of the graphene inclusions needs to be taken into consideration for the best exploitation of graphene as a reinforcing agent in composites. In the middle of the graphene, a remarkably constant strain distribution occurs. Due to the large length of the flake this can be clearly observed even for high level of the applied tension. Moreover, as presented elsewhere[38], the ribbon type geometry of the flake employed here prevents the formation of wrinkling induced by the Poisson contraction of the polymer[38] due to its small width[15].

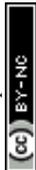

Another crucial point when using graphitic fillers in polymers is the length of the inclusion required for efficient reinforcement of the polymer[18, 39-40]. In general, this length is ten times the transfer length obtained by the shear-lag model, and is estimated as the length required for reaching ~90% of the maximum strain[23] however this analysis is not entirely correct since the transfer length is strain dependent and the level of applied external strain needs to be taken into account. As the results of **figure 3d** clearly show by extrapolation, the monolayer graphene flake can be efficiently loaded at its geometric middle up an external of maximum ~2.8% beyond which it will start lagging behind the applied strain. At that level of strain, if slippage has not initiated, the required critical length (twice the transfer length) is as large as ~60 μm. This value is already much larger than the average commercial flake size[41] and thus it transpires that for a polymer matrix to transfer stress to –say- a simply-supported monolayer graphene a critical length of at least 10 μm is required (**figure 3b**) to achieve strains as low as ~0.5%. This may be quite acceptable for structural composites but it may be a problem for certain functional applications. However it must be stressed that, as has been shown earlier, for fully embedded flakes the transfer lengths have been found to be much smaller than the values reported here for simply-supported flakes[15].

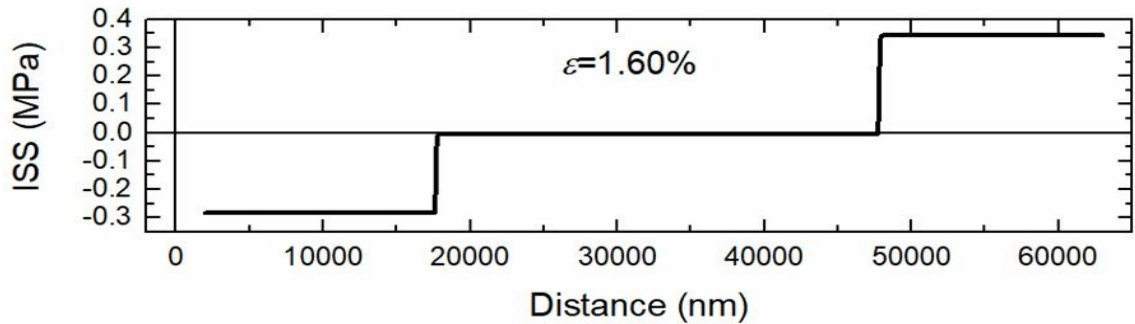

**Figure 4.** The distribution of the ISS across the length of (a) the single layer graphene for ~1.60% of tension. It is clearly observed the increase in the length with the increase in thickness for efficient stress transfer.

We turn our attention now to the stress transfer characteristics of bi and tri-layer graphene-polymer composites. Graphene flakes of large length (>50 μm) and small width were chosen and the corresponding optical images are shown in **figure 2(b, c)**. In this case, almost all edges were fairly square and as will be argued below, shear at the interface was again the main stress transfer mechanism. In the SI we also present results for another bilayer with both edges of angular shape, which shows axial transmittance for both edges and the strain transfer characteristics show certain differences from those presented below. Thus, one can conclude that the edge geometry plays a crucial role in the reinforcing capacity of graphene.

Before we move forward to further discussion of the results of the few-layer flakes, we need to thoroughly consider the validity of equation 2 for this case. The derivation of equation 2 assumes a material of thickness $t$ and Young's modulus $E$ and a particular Raman shift rate. When interlayer slippage occurs equation 2 which is based on load transfer through shear at the graphene/ polymer interface cannot be applied. As the onset of interlayer slippage is hard to detect equation 2 cannot be safely employed in this work to deduct the values of interfacial shear stress. Moreover, it has been found that the adhesion energy of a graphene-substrate interface decreases with the increase in graphene thickness[42], thus we cannot assume that the ISS for the few-layers would be the same with that of the single layer, although this can be tested in future work. It is expected that if the graphene was stretched as a whole unit as for example as in the case of hexagonal boron nitride as shown recently[43], the ISS would be identical for all flakes without layer number dependence where the value of 0.30 MPa was estimated. However, for the current experiments, we can use as a comparative measure of load transfer the slopes of wavenumber shifts per increment of strain as they reflect accurately the stress take-up by the multilayer graphenes.

The average shift rate of the 2D peak is $\sim-43.7\pm2$ cm$^{-1}$/% and $\sim-44\pm2$ cm$^{-1}$/% for the bi and tri-layer flakes, respectively. As pointed out for the case of single layer, the area for the extraction of the shift rate plays a pivotal role. For the bilayer, again the shift is obtained from the central area that exhibits a constant strain profile up to ~1.50%. For the trilayer, we can obtain the shift from the plateau area up to ~1%, but for higher strains the critical length is larger than the lateral flake length so the measurements are inaccurate. These values are slightly smaller than the shift of the single layer but they are in broad agreement with earlier findings[16]. Still, the above values are much higher than those obtained in the past from flakes that were smaller in size than the required critical length for maximum stress transfer[20]. Thus, multi-layers can reach high shift rates having sufficiently large length, but still they are lacking compared to the single layers. Moreover, the good alignment to the loading direction also contributes to this large shift as discussed below. In **figure 5b,e** representative strain profiles for selected levels of tension are introduced in order to present the behavior for the whole examined tensile regime. For both flakes, the results appear similar to those obtained in the case of monolayer graphene albeit at much higher transfer lengths.

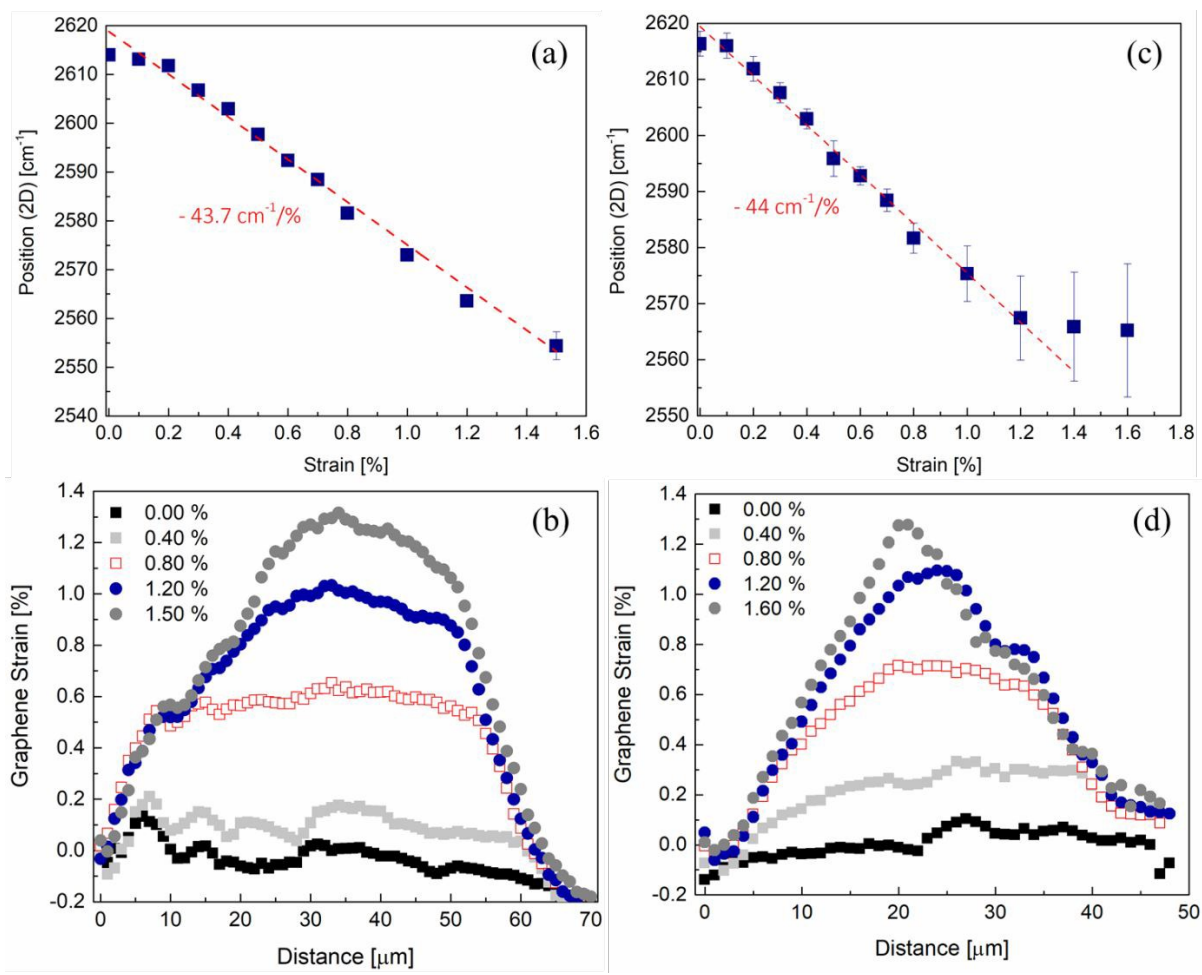

**Figure 5.** The average shift of the 2D peak which has been estimated based on the mid areas to avoid edge effects for (a) bilayer and (c) trilayer, respectively[20]. Representative strain distributions across the length of graphene for selected levels of strain for (b) the bilayer and (d) trilayer flakes, respectively. The open circles represent the ISS obtained before the maximum and constant ISS has been reached.

We observe for the bilayer a convex shape at the right edge which we attribute to its angular shape (**figure 2b**), since the same feature in the strain transfer mechanism was observed for another bilayer with geometrically angular edges (see SI). For the initial levels of strain until ~0.60%, the strain profile fluctuates and becomes smooth as the tension increases. A second build-up occurs at a distance of ~10 microns from the left edge, which is a result of the loss of stacking and interlayer relative slippage which creates a discontinuity in the structure of the multi-layer as observed optically previously[44]. The loss of stacking is clearly demonstrated from the line-shape of the 2D (**figure S6**) peak for both the examined few-layers, and in agreement with previous results[45].

The stress transfer behaviour of the tri-layer is very similar to the bi-layer. Again, we observe problematic build-up for the right edge where axial transmittance occurs (**figure 5e**). For tension of ~0.80% where the strain profile is smooth along the length of the graphene, the shear mechanism is the same friction-like as discussed for the mono and bilayer flakes. Local discontinuities in the strain build-up are also present for the trilayer (**figure 5e**), which is plausibly due to the loss of stacking (see SI for the 2D peak line-shape) as in the case of the bilayer. More results for trilayer flake are given in the SI.

The bilayer of ~70 microns in length can reach critical strain for slippage of ~1.90%, while for the trilayer (see above) that critical point has already been reached at ~1.30% and the interface cannot support any higher stresses/ strains (**figure 6a**). For these strain levels the transfer lengths are of the order of ~15-22 μm for the bilayer which is approximately twice the values obtained for monolayer graphene of square-end at similar strain (**figure 3b**). For the tri-layer the maximum tensile strain has indeed been reached at ~1.30% (**figure 6a**). This is also clearly depicted in the deviation of the 2D shift rate which shows no further downshift of the position of the peak after ~1.20% of tension (**figure 5d**) since the transfer length required for higher tension has overcome the length of the graphene, in contrast with the bilayer which could be stretched to higher strains. A tri-layer of ~50 microns in length can be stretched up to maximum tension of ~1.20% when supported on a polymer while the transfer length at ~1.20% is ~22 to 30 microns, depending on the edge as seen in **figure 6a**. Again, the single layer is beneficial in this regard too, as it needs a much smaller length for reaching the same strain compared to a few-layer. The above results clearly demonstrate that while the same level of interfacial shear or higher can be developed between graphene and polymer for thickness of graphene with one to three layers, the few-layers are still lacking compared to the single layer, due to the inefficient interlayer stress transfer between the individual graphene layers[16, 31]. In cases that graphene needs to reach higher deformations, it is apparent from the previous analysis that graphene flakes with length a few-tens of microns are required, and this length increases for the bilayer and trilayer flakes.

In **figure 6b** the slopes $d\varepsilon/dx$ are plotted versus the number of layers of the graphene for all the examined tensile strain levels. We note that in **figure 6b** the values obtained from both edges for the case of the monolayer are plotted, which show a small decrease for the angular edge compared to the square edge as discussed above. We observe from **figure 6b** that the slopes for the few-layers are strain dependent and increase almost linearly with the tensile strain until a maximum value is reached.  However, the actual values for both bilayer and trilayer are markedly lower than the corresponding shifts of the monolayer for the same strain. This is a direct consequence of the stress transfer process as expressed by the balance-of-forces argument (eq. 2), since for the same interfacial shear stress a thicker material of same stiffness cannot be stressed by the same amount.  The increase of slopes with strain was not expected in view of the onset of interlayer slippage but it may be due to the gradual offset of the compressive residual stresses of the as-prepared specimens. Over approximately 0.5-0.6% of strain the slopes form a plateau but still and in spite of the expected interlayer slippage the system can take up considerable stresses (strains).   Overall, the results obtained here suggest that while the graphene thickness increases much larger lateral sizes are required for efficient reinforcement in the case of commercially produced flakes which exhibit a quite broad distribution of lateral sizes typically of smaller sized than those measured here.

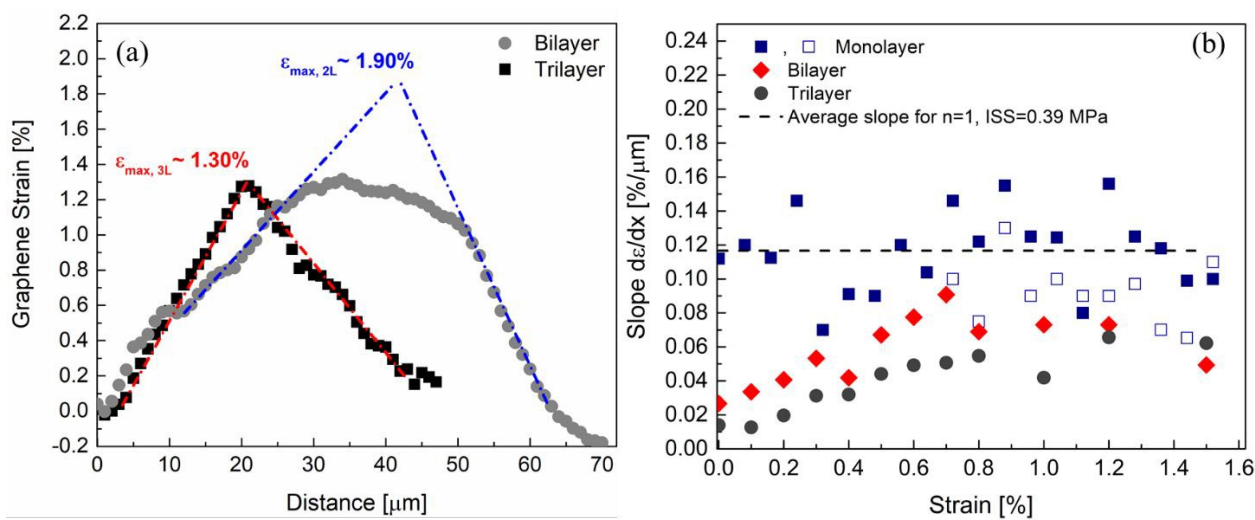

**Figure 6**. (a) Expanding the fitting lines of the strain build-up until convergence for the examined bilayer (grey circles) and trilayer (black rectangles) graphenes. The top value of the formation of the triangle corresponds to the maximum strain that the graphene can reach before slipping from the substrate occurs. The bilayer could reach tensile strain of ~1.90% while the trilayer has already reached its maximum strain. (b) The average values of the slopes obtained from the strain profiles for all flakes and the whole tensile regime. For the single layer the slopes obtained from both edges are plotted.

In summary, we examined the stress transfer mechanism in simply supported graphene polymer-systems with thicknesses of one-to-three layers. The results clearly show that the single layer graphene is more effective as reinforcement agent in polymers in terms of dimensions, since much smaller length is required compared to few-layers for effective reinforcement. We redefined the graphene length required for efficient reinforcement based on the stress transfer through constant ISS which is found to be strain dependent, and estimated that much higher values are actually needed in order to fully exploit the potential of graphene. Graphene flakes with lengths of tens of microns are required for ensuring reinforcing capacity from a polymer substrate to as-supported graphenes at large deformations and the lateral size increases with the increase of graphene thickness. The results also reveal, that squared edges with the applied load are more efficient and lead to higher ISS compared to angular and random shape graphene flakes. The present study provides an in-depth investigation and design guidance when using graphene of random shape and thickness, which are commonly produced by large scale production methods.

## Acknowledgements

The authors acknowledge the financial support of the European Research Council (ERC Advanced Grant 2013) via project no. 321124, "Tailor Graphene. E.N.K. acknowledges receiving funding for this project from the Hellenic Foundation for Research and Innovation (HFRI) and the General Secretariat for Research and Technology (GSRT), under grant agreement No 1536. CG and AM acknowledge support from the Open FET project "Development of continuous two-dimensional defect-free materials by liquid-metal catalytic routes" no. 736299-LMCat which is implemented under the EU-Horizon 2020 Research Executive Agency (REA) and is financially supported by EC.